\documentclass[prl,twocolumn,groupedaddress,showpacs]{revtex4}
\usepackage{bm}

\begin{document}

\title{\textbf{Linear Response for Granular Fluids}}

\author{James Dufty and Aparna Baskaran}
\affiliation{Department of Physics, University of Florida,
Gainesville, FL 32611}

\author{J. Javier Brey}
\affiliation{F\'{\i}sica Te\'{o}rica, Universidad de Sevilla,
Apartado de Correos 1065, E-41080, Sevilla, Spain}

\date{\today }

\begin{abstract}

The linear response of an isolated, homogeneous granular fluid to
small spatial perturbations is studied by methods of
non-equilibrium statistical mechanics. The long wavelength linear
hydrodynamic equations are obtained, with formally exact
expressions for the susceptibilities and transport coefficients.
The latter are given in equivalent Einstein-Helfand and Green-Kubo
forms.  The context of these results and their contrast with
corresponding results for normal fluids are discussed.
\end{abstract}


\maketitle

Granular fluids consist of mesoscopic particles activated by
external forces or initial conditions, and appear ubiquitously
both in nature and in technological applications. Their flow
properties are remarkably similar to those of normal fluids in
many cases \cite{Poschel}. This has led to the speculation that
the methods of non-equilibrium statistical mechanics developed for
normal fluids might be adapted to explore the properties of
granular fluids \cite{Brey97,vanNoije01}. To date, this has been
realized in the examples of numerical molecular dynamics
simulation and kinetic theory. However, the extensive formal
theoretical tools have been applied to a more limited degree
(e.g., impurity diffusion \cite{Dufty02} and liquid state
structure \cite{Lutsko02}). In some respects experimental studies
of the response functions of interest here are more advanced than
those of theory (see, for example, \cite{Xu05}). Here we report an
application of linear response methods to obtain the hydrodynamic
description of small perturbations in an isolated, homogeneous
granular fluid. Although this example is perhaps of limited direct
experimental relevance, it provides an instructive controlled test
for the extension of these methods to granular fluids,
illuminating important differences from normal fluids.

The simplest model of a granular fluid is considered: $N$ smooth,
inelastic, hard spheres ($d=3$) or disks ($d=2$) of mass $m$. The
only difference from a corresponding model for normal fluids is a
loss of energy in each binary collision, characterized by a
restitution coefficient $0<\alpha \leq 1$, with $\alpha =1 $
corresponding to the elastic limit. At the microscopic level, the
fluid state is specified by the positions and velocities of all
particles, i.e. by a point in the $2dN$-dimensional phase space
$\Gamma  \equiv \left\{ {\bm q}_{1}, \cdots ,{\bm q}_{N},{\bm
v}_{1}, \cdots,{\bm v}_{N}\right\} $. This point evolves according
to a deterministic dynamics of free streaming and collisions. The
initial macroscopic state is specified by a probability density
$\rho \left( \Gamma ,0\right)$, characterizing the known
(incomplete) initial data. Consequently, all the ingredients
required for a statistical mechanical description are available.
In particular, the Liouville equation determines the macrostate
$\rho \left( \Gamma ,t\right) $ at all later times. The elements
of a hydrodynamic description then follow from averaging the local
number, energy, and momentum densities to obtain the exact
macroscopic balance equations. From them, the
\emph{phenomenological } hydrodynamic equations are obtained with
the additional assumption of constitutive equations for the
fluxes, as well as for the source term appearing in the equation
for the energy. For systems with small gradients of the
hydrodynamic fields, the Navier-Stokes equations for a granular
fluid are obtained. Of course, the transport coefficients in these
equations are unknown and must be supplied separately from
experiment.

For a normal fluid, a formal theoretical description of
hydrodynamics has been given by considering the linear response of
an equilibrium fluid to small spatial perturbations \cite{Mc89}.
The solutions to the linearized hydrodynamic equations, the
so-called hydrodynamic modes, can then be identified from the long
time, long wavelength behavior of the response functions. As a
result, formally exact expressions for the transport coefficients
are obtained in terms of time correlation functions. These are the
Einstein-Helfand (EH) and Green-Kubo (GK) forms, which have proved
useful as the basis for practical (approximate) theoretical
predictions of these coefficients. The objective here is to report
the application of this linear response approach to granular
fluids.

The isolated granular fluid does not support a stationary,
homogeneous equilibrium state. Instead, the simplest homogeneous
state, known as the homogeneous cooling state (HCS), changes in
time due to collisional energy loss. Spatial perturbations of the
HCS therefore have a dynamics due to both the perturbation and the
background reference state. However, the latter occurs only
through the average energy and can be removed by a suitable choice
of dimensionless variables. Effectively then, the analysis is
similar to that for the equilibrium case, although the reference
state is not the Gibbs distribution. The distribution function of
the HCS, $\rho_{0}(\Gamma)$, is an example of a ``normal
distribution'', in the sense that its time dependence occurs
entirely through the temperature,
\begin{equation}
\rho _{0}(\Gamma,t) = (\ell v_{0}(t))^{-dN} \rho _{0}^{\ast}
\left( \{  {\bm  q}_{ij}/\ell, {\bm v}_{i}/v_{0}(t)
\},n_{0}\right). \label{1}
\end{equation}
Here ${\bm q}_{ij} \equiv {\bm q}_{i}-{\bm q}_{j}$  and $\ell $ is
the mean free path. The uniform hydrodynamic fields are the
average number density, $n_{0}$, and the temperature, $T_{0}(t)$,
the latter occurring through the thermal velocity $v_{0}(t) \equiv
\sqrt{2T_{0}(t)/m}$ . The granular temperature is defined in terms
of the average energy density, $ e_{0}$, in the same way as for
equilibrium, $e_{0}=dn_{0}T_{0}/2$. The HCS distribution is the
solution to the Liouville equation in the form \cite{Dufty02}
\begin{equation}
\overline{\mathcal{L}}\rho _{0}=0, \quad
\overline{\mathcal{L}}\equiv \overline{L}+\frac{\zeta_{0}}{2}
\sum_{i=1}^{N}\left( d+{\bm v}_{i}\cdot \frac{\partial}{\partial
{\bm v}_{i}} \right) , \label{2}
\end{equation}
where $\overline{L}$ is the Liouville operator for hard spheres
\cite{Brey97,vanNoije01}, and $\zeta _{0} \equiv -\partial _{t}\ln
T_{0}$ is the ``cooling rate''.

To study the response of the HCS to spatial perturbations, an
initial {\em local} HCS, $\rho _{l}(\Gamma,0)$, is chosen. This
distribution is a functional of specified nonuniform hydrodynamic
fields $n\left( {\bm r}\right)$, $ T\left( {\bm r}\right)$, and
${\bm u}\left( {\bm r}\right)$, characterizing the average number
density, energy density, and momentum density for this ensemble.
To construct it, consider first the HCS for an inhomogeneous fluid
(i.e., the solution to (\ref{2}) in an external inhomogeneous
field). Since the average nonuniform density is a functional of
the external field, inverting that relationship gives the HCS
distribution as a functional of the density $\rho _{0}^{(inh)}=
(\ell v_{0})^{-dN}\rho _{l}^{\ast } \left( \{ {\bm
q}_{ij}/\ell,{\bm v}_{i}/v_{0}\} | n\right) $. The special case in
(\ref{1}) is the value of this functional at the uniform state.
Next, the non-uniform flow field and temperature are incorporated
through the replacements of ${\bm v}_{i}$ by ${\bm v}_{i}-{\bm
u}({\bm q}_{i})$ and $v_{0}(T_{0})$ by $v_{0}({\bm q}_{i}) \equiv
v_{0}[T({\bm q}_{i})]$ for each particle in the arguments of
$\rho_{0}^{(inh)}$. The local HCS distribution is then
\begin{equation}
\rho _{l}(\Gamma) \equiv \left\{ \sum_{i=1}^{N} \left[\ell
v_{0}({\bm q}_{i})\right]^{-d} \right\} \rho _{l}^{\ast } \left(
\left\{ \frac{{\bm q}_{ij}}{\ell},\frac{{\bm v}_{i}- {\bm u} (
{\bm q}_{i}) }{v_{0}({\bm q}_{i})} \right\} | n \right) .
\label{3}
\end{equation}
This initial preparation is natural, in the sense that it is
expected that locally there is a rapid relaxation of the velocity
distribution as each region attempts to approach the HCS at its
own values of the hydrodynamic fields. It is the analogue of rapid
collisional approach to a local equilibrium state for normal
fluids. For smooth spatial variations of small amplitude, to
linear order
\begin{eqnarray}
\rho (0) &=&\rho _{l}(0) \simeq \rho _{0}+ \sum_{\beta} \int d{\bm
r}\, \left[ \frac{\delta \rho _{l}(0)}{\delta y_{\beta }({\bm
r},0)}\right]_{0} \delta
y_{\beta }({\bm r},0)  \nonumber \\
&=&\rho _{0}(\Gamma,0)+\sum_{\beta} \phi _{\beta }(-{\bm k})\delta
\widetilde{y}_{\beta }({\bm k},0), \label{4}
\end{eqnarray}
where the $y_{\beta }(\mathbf{r},0)$ are linear combinations of
the hydrodynamic fields, as specified below. The second equality
follows by considering a single Fourier component for these
fields, and it defines $\phi_{\beta}({\bm k})$.

This preparation is also special in the sense that it excites only
hydrodynamic modes in the long wavelength limit. This follows from
the fact that the $y_{\beta }$ can be chosen such that
$\Phi_{\beta} \equiv \phi_{\beta}({\bm k}=0)$ are eigenfunctions
of the extended Liouville operator $\overline{\mathcal{L}}$
\cite{Dufty05}
\begin{equation}
\overline{\mathcal{L}}\Phi _{\beta }=\lambda _{\beta }\Phi _{\beta
}, \quad \Phi _{\beta }\equiv \int d{\bm r}\, \left[ \frac{\delta
\rho _{l}}{ \delta y_{\beta }(\mathbf{r},0)}\right]_{0}, \label{5}
\end{equation}
with the eigenvalues $\{ \lambda_{\beta} \} = \{
0,\zeta_{0}/2,-\zeta_{0}/2,..,-\zeta_{0}/2 \}$. The specific set
of fields in (\ref{4}) leading to this result are
\begin{equation}
\{ y_{\beta} \} = \left\{ \frac{ n}{n_{0}}\, ,\frac{ T
}{T_{0}}+2\, \frac{\partial \ln \zeta _{0}}{\partial \ln
n_{0}}\frac{n}{ n_{0}}\, ,\frac{\widehat{\bm k}\cdot {\bm
u}}{v_{0}}\, , \frac{{\bm u}_{\perp}}{v_{0}} \right\}. \label{7}
\end{equation}
The field flow ${\bm u}$ has been decomposed into its longitudinal
component in the direction of $\widehat{\bm k}={\bm k}/k$, and the
remaining $d-1$ transversal components denoted by ${\bm
u}_{\perp}$. The eigenvalue $-\zeta_{0}/2$ is $d$-fold
degenerated. Significantly, the eigenvalues $\lambda_{\beta}$ are
the same as those of the linearized macroscopic balance equations
for these variables in the long wavelength limit. Hence they can
be considered as hydrodynamic modes reflected in the microscopic
dynamics. These observations are critical to the following linear
response analysis. In the elastic limit, the eigenfunctions become
linear combinations of the global conserved number, energy, and
momentum.

The response of the hydrodynamic fields $\delta y_{\beta }$ to the
initial perturbation (\ref{4}) is
\begin{equation}
\delta \widetilde{y}_{\beta }^{*}({\bm k}^{*},s)= \sum_{\gamma}
C_{\beta \gamma }({\bm k^{*}} ,s)\delta \widetilde{y}_{\gamma
}^{*}({\bm k}^{*},0), \quad  ds=\frac{v_{0}}{\ell}\, dt, \label{8}
\end{equation}
\begin{equation}
C_{\beta \gamma }({\bm k}^{*},s) \equiv \int d\Gamma^{*} \delta
Y_{\beta }^{*}({\bm k}^{*}) e^{-s \overline{\mathcal{L}}^{*}}\phi
_{\gamma }^{*}(-{\bm k}). \label{9}
\end{equation}
The dimensionless time $s$ is a measure of the average number of
collisions, and $Y_{\beta }$ is the phase function whose average
is $y_{\beta }$. Moreover, the asterisks indicate dimensionless
variables in the units defined by $\ell$, $v_{0}$, and $m$ \cite
{Dufty02}. Clearly, if the $\delta \widetilde{y}_{\beta }^{*}({\bm
k}^{*},s)$ obey hydrodynamic equations at long space and time
scales, the corresponding excitations must exist in the spectrum
of the correlation functions $ C_{\beta \gamma }({\bm k}^{*},s)$.
This suggests defining an exact transport matrix
$\mathcal{K}_{\alpha \beta }({\bm k}^{*},s)$ by
\begin{equation}
\left[ \partial _{s}+\mathcal{K}({\bm k}^{*},s)\right] \delta
\widetilde{y}^{*}({\bm k}^{*},s)=0,  \label{10}
\end{equation}
with the identification
\begin{equation}
\mathcal{K}({\bm k}^{*},s)=-\left[ \partial _{s}C({\bm
k}^{*},s)\right] C^{-1}({\bm k}^{*},s).  \label{11}
\end{equation}
A matrix notation has been introduced for simplicity. The
hydrodynamic equations are then identified from the matrix
$\mathcal{K}({\bm k}^{*},s)$ for large $s$ and small ${\bm
k}^{*}$. In particular, the linear Navier-Stokes approximation is
defined by an expansion of $\mathcal{ K}({\bm k}^{*},s)$ to order
$k^{*2}$, with the coefficients evaluated for large $s$. In
general, this procedure does not lead to tractable results since
initial perturbations excite microscopic as well as hydrodynamic
modes. However, the special initial preparation chosen here is
such that only hydrodynamic excitations occur in the long
wavelength limit. This follows directly from (\ref{5}) and
(\ref{9})
\begin{equation}
C_{\beta \gamma }(0,s)=e^{-s \lambda^{*} _{\beta }}\delta _{\beta
\gamma }.  \label{12}
\end{equation}
As a consequence, the expansion of (\ref{11}) is assured to start
on the hydrodynamic branch. Define
\begin{equation}
C\left( {\bm k}^{\ast },s\right) =e^{- s \Lambda }+ik^{*}
C^{(1)}\left( s\right) +\left( ik^{*}\right) ^{2}C^{(2)}\left(
s\right) + \cdots, \label{13}
\end{equation}
where $\Lambda _{\alpha \beta }=\lambda^{*}_{\alpha }\delta
_{\alpha \beta }$. Then it is found
\begin{equation}
\mathcal{K}({\bm k}^{\ast },s ) =\Lambda +ik^{*} \mathcal{K}
^{(1)}(s) + ( ik^{*})^{2}\mathcal{K}^{(2)}(s) + \cdots, \label{14}
\end{equation}
with
\begin{equation}
\mathcal{K}^{(1)}(s) = - \left[ ( \partial _{s}+\Lambda
)C^{(1)}(s) \right] e^{s \Lambda },  \label{15}
\end{equation}
\begin{equation}
\mathcal{K}^{(2)}(s) =-\left[ ( \partial_{s}+\Lambda )C^{(2)}(s)
+\mathcal{K}^{(1)}(s) C^{(1)}(s) \right] e^{s\Lambda }\, .
\label{16}
\end{equation}
The transport coefficients and susceptibilities of the
phenomenological Navier-Stokes equations can be identified with
appropriate matrix elements in these expressions:
$\mathcal{K}^{(1)}$ gives the Euler coefficients,
$\mathcal{K}^{(2)}$ gives the Navier-Stokes coefficients, and so
on. In particular, these expressions provide an extension of the
EH representation of transport coefficients to granular fluids.
The GK forms follow from these results by partial integration. The
detailed expressions and analysis will be given elsewhere.

In this brief presentation, further details will be limited to the
case of the transverse flow field response, i.e. $\delta
\widetilde{ y}_{\perp}^{*}({\bm k}^{*},s)$ associated with one
component of ${\bm u}_{\perp}$ in (\ref{7}). The only
non-vanishing element $C_{\alpha \perp}$ is $C_{\perp \perp}$.
Also note that $C_{\perp \perp}^{(1)}=0$ from symmetry. Then, Eq.\
(\ref{10}) becomes to order $k^{2}$
\begin{equation}
\left[ \partial _{s}- \frac{\zeta^{*}_{0}}{2}-
k^{*2}\mathcal{K}_{\perp \perp}^{(2)}(s)\right] \delta \widetilde{
y}_{\perp}^{*}({\bm k}^{*},s)=0.  \label{17a}
\end{equation}
If $\mathcal{K}_{\perp \perp}^{(2)}(s)$ has a limit for large $s$,
this becomes the shear diffusion equation, with
$\mathcal{K}_{\perp \perp}^{(2)}$ identified in terms of the
dimensionless shear viscosity $\eta^{*}\equiv \eta/m n_{0} \ell
v_{0}$,
\begin{eqnarray}
\eta^{*} &=&-\lim_{s>>1}\mathcal{K}_{\perp
\perp}^{(2)}(s)=\lim_{s>>1}\partial _{s}\left[ C_{\perp \perp
}^{(2)}(s) e^{-\frac{s \zeta_{0}^{*}}{2}}\right]
\nonumber \\
&=&\lim_{s>>1}\partial _{s}\frac{1}{N}\int d\Gamma^{*} \left[ e^{s
\left( \mathcal{L}^{*}-\frac{\zeta^{*}_{0}}{2}\right)} M^{*}
\right] \mathcal{N}^{*}\rho_{0}^{*},  \label{17b}
\end{eqnarray}
where $-\mathcal{L}^{*}$ is the adjoint of
$\overline{\mathcal{L}}^{*}$. The phase function $M^{*}$  is the
space moment of the momentum density
\begin{equation}
M^{*}=\sum_{i=1}^{N} (\widehat{\bm k} \cdot {\bm q}^{*}_{i}) (
\widehat{\bm e}_{1} \cdot {\bm v}^{*}_{i}) , \label{19}
\end{equation}
where $\widehat{\bm e}_{1}$ is the unit vector in the direction of
${\bm u}_{\perp}$ considered. The function $\mathcal{N}^{*}
\rho_{0}^{*}\equiv \phi _{\perp}^{(1)}$ is given by
\begin{eqnarray}
\mathcal{N}^{*}\rho_{0}^{*}&=& -\int d{\bm r}^{*}\, \left(
\widehat{\bm k} \cdot {\bm r}^{*} \right) \widehat{\bm e}_{1}\cdot
\left[ \frac{\delta \rho^{*}_{l }}{\delta {\bm u}^{*}({\bm
r},0)}\right]_{0} \nonumber \\
&=& \sum_{i=1}^{N}\left( \widehat{\bm k}\cdot {\bm
q}_{i}^{*}\right) \widehat{\bm e}_{1}\cdot
\frac{\partial}{\partial {\bm v}_{i}^{*}}\, \rho _{0}^{*}.
\label{20}
\end{eqnarray}
In the elastic limit and for $\rho _{0} \rightarrow $ $\rho _{e}$,
the equilibrium canonical distribution function, the Helfand form
for the shear viscosity of a normal fluid is recovered
\begin{equation}
\eta^{*}_{elast} =- \lim_{s>>1}\partial _{s}\frac{2}{N}\int
d\Gamma^{*} \left( e^{s L^{*}}M^{*}\right) M^{*} \rho^{*}_{e}.
\label{21}
\end{equation}
Both (\ref{17b}) and (\ref{21}) contain correlation functions
involving the momentum density moment $M$. For a normal fluid.
this is an autocorrelation function, while for the granular fluid
a new moment $\mathcal{N}$ appears that is determined from the HCS
distribution.

These correlation functions must grow as $s$ for long times to
obtain a constant viscosity. This is analogous to the definition
of the diffusion coefficient for a Brownian particle in terms of
its mean square displacement, and for that reason the
representation above is referred to as the EH form. An equivalent
representation is obtained by carrying out the time derivative,
shifting the time dependence, and integrating with respect to
time. The result is
\begin{eqnarray}
\eta^{*} & = & \frac{1}{N}\int d\Gamma^{*}\,  F^{*}\mathcal{N}^{*}
\rho^{*}_{0} \nonumber \\
& + & \lim_{s \ll 1}\int_{0}^{s}ds^{\prime }\frac{1}{N}\int
d\Gamma^{*} F^{*}e^{- s^{\prime} \left(
\overline{\mathcal{L}}^{*}+\frac{\zeta_{0}^{*}}{2}\right)}
\mathcal{F}^{*}\rho_{0}^{*}.  \label{22}
\end{eqnarray}
The fluxes are given by
\begin{equation}
F^{*}=\left( \mathcal{L}^{*}-\frac{\zeta_{0}^{*}}{2}\right) M^{*},
\quad \mathcal{F}^{*} \rho_{0}^{*}=-\left(
\overline{\mathcal{L}}^{*}+\frac{\zeta_{0}^{*}}{2} \right)
\mathcal{N}^{*} \rho_{0}^{*}.  \label{23}
\end{equation}
This is the GK representation. For normal fluids, $F^{*}$ becomes
the volume integrated momentum flux and the second term of
(\ref{22}) becomes the expected time integral of a flux
autocorrelation function. As noted below, hard sphere systems
entail the additional first term on the right side even for normal
fluids. Moreover, as with the EH form, the correlation functions
for the granular fluid involve new phase functions defined from
the HCS distribution.

The objective here has been to explore the application of linear
response methods to the simplest model and state of a granular
fluid. The details for all hydrodynamic modes and all transport
coefficients will be reported elsewhere. These results illustrate
the applicability of such methods and their potential for more
complex conditions of experimental interest. They also expose many
subtleties in the differences from corresponding results for
normal fluids and provide a warning against a simple translation
of normal fluid theory to granular fluids. In closing, some
observations are highlighted:

i) The identification of the transport matrices in
(\ref{14})-(\ref{16})  constitutes a formal derivation of the
linear Navier-Stokes equations. Important limitations on this
derivation are placed by restrictions on length and time scales.
As for normal fluids, it is expected that the conditions are times
long compared to the mean free time ($s \gg 1$) and wavelengths
long compared to the mean free path ($k^{*}=k \ell \ll 1$).

ii) These formal results have no \emph{a priori} limitation on the
density or degree of inelasticity. However, peculiarities of the
isolated system (cluster instability \cite{GyZ93}, inelastic
collapse \cite{McyY96}) must be confronted at high densities and
large inelasticity. These problems can be removed in practice by
considering small system sizes and protocols for simulations to
avoid dangerous configurations with no physical consequences.

iii) The transport coefficients are given in terms of averages
over the HCS reference state, which differs from the equilibrium
canonical distribution. Consequently, one of the two phase
functions in the EH and GK expressions is different from that for
normal fluids.

iv) The generator $\overline{L}$ for hard sphere dynamics is
singular and must be expressed in terms of binary collision
operators rather than forces. Important structural changes result
in the form of the GK expressions and the associated microscopic
fluxes \cite{Dufty022}. The GK expressions for both elastic and
inelastic hard sphere fluids have a new contribution due to
instantaneous collisions, as well as the usual time integral of
correlation functions of fluxes. The latter are expressed in terms
of the binary collision operators and are different for forward
and backward dynamics, both occurring in the GK expressions.

v) The reference HCS is time dependent, due to collisional
cooling. A dimensionless form of the equations is required to
eliminate this dependence, resulting in a new generator
$\overline{\mathcal{L}}^{*}$ rather than the Liouville operator
$\overline{L}$. In this representation, the HCS is a stationary
solution, $\overline{\mathcal{L}}^{*} \rho _{0}^{*}=0 $.

vi) The stationary HCS distribution is singular in the sense that
it is restricted to a zero total momentum ${\bm P}$ surface, i.e.,
$\rho _{0}^{*} \propto \delta \left( {\bm P}^{*}\right) $.
Furthermore, the elastic limit is proportional to a delta function
in the total energy, since an isolated system is being considered.
This constant momentum, constant energy ensemble is often referred
to as the ``computer'' ensemble, as it corresponds to usual
conditions in molecular dynamics simulation. For granular fluids,
the energy changes, even in isolation, but the constant momentum
condition must be retained for consistency with
$\overline{\mathcal{L}}^{*} \rho _{0}^{*}=0$.

vii) The presence of hydrodynamic excitations is expected for a
wide class of initial perturbations. However, the choice made here
is both physically and mathematically motivated. As the HCS
distribution characterizes the homogeneous state through its
dependence on the hydrodynamic fields, it is natural to consider
their spatial perturbation through this same dependence. More
important for the analysis here is that the resulting perturbation
excites \emph{only} hydrodynamic modes in the long wavelength
limit. This follows from the property (\ref{5}), and plays a
central role in deriving the explicit forms for the expansion
(\ref{14}). For general initial preparations, the leading term in
(\ref{14}) will not be $\Lambda $, but rather some complex time
dependent quantity that must be calculated to be $\Lambda $ at
long times.

viii) An earlier derivation of GK expressions leads to different
results \cite{Goldhirsh00}. The method is formally correct, but
does not have the simplifying features of the initial preparation
considered here. In addition, some of the special properties of
the hard sphere generator were not accounted for in detail, so
that different fluxes are obtained and the instantaneous
contribution is missed.

The research of J.D. and A.B. was supported in part by the
Department of Energy Grant (DE-FG03-98DP00218). J.J.B.
acknowledges partial support from the Ministerio de Educaci\'{o}n
y Ciencia (Spain) through Grant No. BFM2005-01398 (partially
financed by FEDER funds). This work also was supported in part by
the National Science Foundation under Grant No. PHY99-0794 to the
Kavli Institute for Theoretical Physics, UC Santa Barbara.

\end{document}